\documentclass{article}
\usepackage{graphicx}
\usepackage{amsmath}
\usepackage{amsfonts}
\usepackage{amssymb}
\newtheorem{theorem}{Theorem}

\newtheorem{lemma}[theorem]{Lemma}

\newtheorem{proposition}[theorem]{Proposition}

\newenvironment{proof}[1][Proof]{\textbf{#1.} }{\ \rule{0.5em}{0.5em}}

\begin{document}

\title{ \textbf{Random Lattices and Random Sphere Packings: Typical Properties}}
\author{Senya Shlosman\\Centre de Physique Theorique, \\CNRS, Luminy, case 907, F-13288, \\Marseille Cedex 9, France, and \\IPPI, RAS, Moscow, Russia\\\textit{shlosman@cpt.univ-mrs.fr}
\and Michael A. Tsfasman\\IPPI, RAS, Moscow, Russia,\\Independent University of Moscow, and\\IML-CNRS, Luminy. \\\textit{tsfasman@iitp.ru}}
\maketitle
\begin{abstract}
We review results about the density of typical lattices in $\mathbb{R}^{n}.$
They state that such density is of the order of $2^{-n}.$ We then obtain
similar results for random packings in $\mathbb{R}^{n}:$ after taking suitably
a fraction $\nu$ of a typical random packing $\sigma$, the resulting packing
$\tau$ has density $C\left(  \nu\right)  2^{-n},$ with a reasonable $C\left(
\nu\right)  .$ We obtain estimates on $C\left(  \nu\right)  .$
\end{abstract}

\section{Introduction}

The problem of filling the euclidean space $\mathbb{R}^{n}$ with equal
non-overlapping spheres has a long and celebrated history. It was started by
J.Kepler in 1610, who studied the hexagonal two-dimensional packing, and
conjectured that the face-centered cubic three-dimensional one is the densest
possible. Later on, I.Newton claimed that in $\mathbb{R}^{3}$ there cannot be
more than 12 non-intersecting balls of unit radius touching a given unit ball.
In the case when the centers of spheres form a lattice (i.e., a discrete
additive subgroup of $\mathbb{R}^{n}$) the question of finding lattices with
densities high enough was studied by C.F.Gauss. Starting from dimension $9,$
the highest density of the lattice packing is still unknown. Naturally, even
less is known when the ball arrangement is not of a lattice nature. The
problem constitutes an essential part of 18-th Hilbert problem. Whether or not
the best density is achieved on the lattice arrangements is a major open
question in the field. Quite recently T.C.Hales and S.P.Ferguson seem to have
proved that for $n=3$ the largest density is that of the face-centered cubic
lattice packing . The proof is very long and computer assisted. It is
discussed in \cite{Oe2}. Constructions of dense enough sphere packings are
often based on rather subtle algebraic technique, sometimes using algebraic
geometry and number theory methods (cf. \cite{CS}, \cite{TV}, \cite{RT},
\cite{Oe1}).

In this paper we are not trying to solve any of the above problems. Instead,
we address the question of the density of \textit{typical} lattice packings
and of \textit{typical} random packings. Here, the word ``typical'' refers to
the natural probability distributions, defined below. In essence, the
probabilistic approach to the lattice packing problem is not new, and existing
results on dense enough lattice packings are implicitly based on it (cf.
\cite{M,H,Sch}). So, the results about the typical lattices presented below,
are rather of the review nature, though they are presented in a new way. The
most important element in them is probably the concept of a typical lattice
itself. The importance of these results for us is that they give the reference
frame we need when we discuss the properties of random packings, the main
topic of the present paper.

As it is usual in the probability theory, the reasonable results can be
obtained only in the limit when the number of degrees of freedom of the system
goes to infinity. For the case of lattices that means passing to the limit
over the dimension $n\rightarrow\infty.$ On the other hand, for random
packings we have infinitely many degrees of freedom already in the (infinite
volume) finite dimensional euclidean space $\mathbb{R}^{n},$ so no other limit
is needed here.

Below we fix some notation. In the next section we present our results
concerning the properties of the typical lattice packings. The last section
deals with the random packings.\medskip

Let $\sigma$ be a locally finite subset of $\mathbb{R}^{n}.$ Let $V_{N}%
\subset\mathbb{R}^{n}$ denote a cube with the side $2N,$ centered at the
origin. By $\sigma_{N}$ we denote the intersection $\sigma\cap V_{N}.$ Let
$d\left(  \sigma_{N}\right)  $ denote the minimal spacing between the points
of $\sigma_{N}:$%

\[
d\left(  \sigma_{N}\right)  =\min_{x,y\in\sigma_{N}:x\neq y}\left|
x-y\right|  ,
\]
while $r\left(  \sigma_{N}\right)  =\dfrac12d\left(  \sigma_{N}\right)  .$ We
define the \textit{geometric (or sphere packing) density} of the configuration
$\sigma$ in the box $V_{N}$ to be the number
\[
\Delta_{N}\left(  \sigma\right)  =\dfrac{\mathrm{vol}\left[  \left(
\bigcup_{x\in\sigma_{N}}B\left(  x,r\left(  \sigma_{N}\right)  \right)
\right)  \bigcap V_{N}\right]  }{\left(  2N\right)  ^{n}}.
\]
Here $B\left(  x,r\right)  $ is a ball of radius $r$ centered at the point $x.$

The \textit{sphere packing density} of an infinite volume configuration
$\sigma$ is defined as%

\[
\Delta\left(  \sigma\right)  =\limsup_{N\rightarrow\infty}\Delta_{N}\left(
\sigma\right)  .
\]
This corresponds to the classical problem of packing equal non-overlapping
spheres in $\mathbb{R}^{n}.$

In the special case of \textit{lattice packings }the set $\sigma$ is a
discrete additive subgroup of $\mathbb{R}^{n}$ and the density can be
rewritten as
\[
\Delta\left(  \sigma\right)  =\frac{v_{n}r\left(  \sigma\right)  ^{n}}%
{\det\left(  \sigma\right)  },
\]
where $v_{n}\left(  r\right)  =r^{n}v_{n}\left(  1\right)  $ is the volume of
the sphere of the radius $r$ in $\mathbb{R}^{n},$ $v_{n}=v_{n}\left(
1\right)  ,$ and $\det\left(  \sigma\right)  $ is the volume of the
fundamental domain of $\sigma$.

The classical question then is about the largest possible value of
$\Delta\left(  \sigma\right)  $ for any locally finite subset $\sigma
\subset\mathbb{R}^{n}$ or for any lattice $\sigma\subset\mathbb{R}^{n},$ and
also how to construct the corresponding $\sigma$-s.

In this paper we discuss a different problem. We are interested in the density
of a typical (random) configuration $\sigma,$ and that of a typical (random)
lattice. Thus, our main results deal with the distribution of the random
quantity $\Delta\left(  \sigma\right)  $. Namely, we show that in the lattice
case the geometric density of a typical lattice is of the order of $2^{-n}$
(see Theorem \ref{A} below). For the random sphere packing the geometric
density of a typical realization of a corresponding random field is zero. The
problem becomes interesting if we allow ourselves to decimate the
configuration and to throw away a fraction of it, containing `bad' sites. Then
it turns out that the remaining random configuration has the geometric density
of the same order as in the lattice case (see Theorem \ref{C} below), for a
proper choice of the bad set.

\section{Random lattices}

In this section we discuss the lattice case. In the first subsection we
present the results concerning the case of very high dimension $n,$ which turn
into a simple relation in the limit $n\rightarrow\infty.$ In the following
subsection we consider the case $n=2,$ when again the results can be expressed
by simple relations, due to explicit computations.

\subsection{The case of large dimensions}

Let $\sigma$ is a lattice. Clearly, its geometric density remains the same
after multiplication by a scalar. Therefore, we can consider here only
unimodular lattices, i.e., those with $\det\left(  \sigma\right)  =1$ . The
space of unimodular lattices in $\mathbb{R}^{n}$ is naturally isomorphic to
the symmetric space $\Lambda_{n}=SL_{n}\left(  \mathbb{R}\right)
/SL_{n}\left(  \mathbb{Z}\right)  $. It is equipped with the Haar measure
$\mu_{n}$ and $\mu_{n}\left(  \Lambda_{n}\right)  $ is finite. We therefore
can normalize it so that $\mu_{n}\left(  \Lambda_{n}\right)  =1$. After such
choice the density $\Delta\left(  \sigma\right)  $ becomes a random variable.
As the following theorem shows, in order to get a nontrivial limiting
distribution for it, one should normalize it by the factor $2^{n}:$

\begin{theorem}
\label{A} Let
\[
F_{n}\left(  x\right)  =\mu_{n}\left\{  \sigma\in\Lambda_{n}:2^{n}%
\Delta\left(  \sigma\right)  \leq x\right\}
\]
be the distribution function of the random variable $2^{n}\Delta\left(
\sigma\right)  $ . Then
\[
\lim_{n\rightarrow\infty}F_{n}\left(  x\right)  =1-e^{-x/2}.
\]
\end{theorem}

This result follows from the following theorem of Schmidt:

\begin{theorem}
\label{B}\cite{Sch} Let $n\geq13.$ Let $S$ be a Borel set in $\mathbb{R}^{n}$
such that $S\cap\left(  -S\right)  =\emptyset.$ Suppose that $\mathrm{vol}%
\left(  S\right)  \leq n-1$ . Then the measure
\[
\mu_{n}\left\{  \sigma\in\Lambda_{n}:\sigma\cap S=\emptyset\right\}
=e^{-\mathrm{vol}\left(  S\right)  }\left(  1-R_{n}\right)  ,
\]
where
\[
\left|  R_{n}\right|  <6\left(  \frac{3}{4}\right)  ^{n/2}e^{4\mathrm{vol}%
\left(  S\right)  }+\mathrm{vol}\left(  S\right)  ^{n-1}n^{-n+1}%
e^{\mathrm{vol}\left(  S\right)  +n}.
\]%
\endproof
\end{theorem}

\textbf{Proof of Theorem \ref{A}.} To derive Theorem \ref{A} from Theorem
\ref{B} take $S=S\left(  d\right)  $ to be ``one half of the ball'' $B\left(
0,d\right)  $, in such a way as to satisfy the conditions $S\cap\left(
-S\right)  =\emptyset$ and $S\cup\left(  -S\right)  =B\left(  0,d\right)
\,\backslash\,0.$ Then, for a lattice $\sigma$ the condition $\sigma\cap
S=\emptyset$ is equivalent to the condition that $d\left(  \sigma\right)  >d,
$ so
\begin{align*}
\mu_{n}\left\{  \sigma\in\Lambda_{n}:d\left(  \sigma\right)  >d\ \right\}   &
=\mu_{n}\left\{  \sigma\in\Lambda_{n}:\sigma\cap S\left(  d\right)
=\emptyset\right\} \\
&  =e^{-\mathrm{vol}\left(  S\left(  d\right)  \right)  }\left(
1-R_{n}\right)  .
\end{align*}
Let us put $x\left(  \sigma\right)  =v_{n}d\left(  \sigma\right)  ^{n},$ then
$2^{n}\Delta\left(  \sigma\right)  =x\left(  \sigma\right)  ,$ since
$\det\left(  \sigma\right)  =1.$ Put $d=\left(  \frac x{v_{n}}\right)
^{1/n}.$ Then, by definition
\begin{align*}
F_{n}\left(  x\right)   &  =\mu_{n}\left\{  \sigma\in\Lambda_{n}:2^{n}%
\Delta\left(  \sigma\right)  \leq x\right\} \\
&  =\mu_{n}\left\{  \sigma\in\Lambda_{n}:x\left(  \sigma\right)  \leq
x\right\} \\
&  =1-\mu_{n}\left\{  \sigma\in\Lambda_{n}:d\left(  \sigma\right)  >d\right\}
\\
&  =1-\mu_{n}\left\{  \sigma\in\Lambda_{n}:\sigma\cap S\left(  d\right)
=\emptyset\right\}  .
\end{align*}
Noting that $\mathrm{vol}\left(  S\left(  d\right)  \right)  =\frac
x2,\mathrm{\ }$we have
\begin{align*}
F_{n}\left(  x\right)   &  =1-\mu_{n}\left\{  \sigma\in\Lambda_{n}:\sigma\cap
S\left(  d\right)  =\emptyset\right\}  =1-e^{-\mathrm{vol}\left(  S\left(
d\right)  \right)  }\left(  1-R_{n,x}\right) \\
\  &  =1-e^{-x/2}\left(  1-R_{n,x}\right)  ,
\end{align*}
with
\[
\left|  R_{n,x}\right|  <6\left(  \frac34\right)  ^{n/2}e^{4x/2}+\left(  \frac
x2\right)  ^{n-1}n^{-n+1}e^{x/2+n}\rightarrow0
\]
as $n\rightarrow\infty$, which proves our statement.%

\endproof

\subsection{Two-dimensional case}

For lattices in $\mathbb{R}^{2}$ the information is much more precise.

Consider lattice packings in $\mathbb{R}^{2}$, i.e., lattices in $\mathbb{C}$.
Instead of fixing their determinant, let us use the freedom of scalar
multiplication and rotation to fix the shortest basis vector to be $1$ . Then
we can take the other basis vector $z$ in the modular domain
\[
F=\left\{  x^{2}+y^{2}\ge1,-\frac12\le x\le\frac12\right\}  .
\]
The corresponding lattice will be denoted by $L_{z}.$ Its packing radius is
$\frac12.$ If $z=x+iy,$ then the volume of the fundamental parallelepiped of
$L_{z}$ is $y.$ Therefore, its sphere packing density $\Delta\left(
L_{z}\right)  $ equals $\frac\pi{4y}$.

The space of two-dimensional lattices $\Lambda_{2}=SL_{2}\left(
\mathbb{R}\right)  /SL_{2}\left(  \mathbb{Z}\right)  $ is canonically
homeomorphic to what we get from $F$ by identifying every boundary point
$\left(  x,y\right)  $ with the boundary point $\left(  -x,y\right)  .$

We will use the well-known fact that the probability Haar measure on
$\Lambda_{2}$ in the $\left(  x,y\right)  $ coordinates is given by
$d\mu=\frac3\pi\frac{dxdy}{y^{2}}.\,$ The sphere packing density $\Delta$ then
becomes a random variable.

\begin{theorem}
The random variable $\Delta$ has a density $p_{\Delta}\left(  x\right)  $,
given by
\[
p_{\Delta}\left(  x\right)  =\left\{
\begin{array}
[c]{cc}%
\frac{12}{\pi^{2}} & \text{for }0\leq x\leq\frac{\pi}{4},\\
\frac{12}{\pi^{2}}\left(  1-2\sqrt{1-\left(  \frac{\pi}{4x}\right)  ^{2}%
}\right)  & \text{for }\frac{\pi}{4}\leq x\leq\frac{\pi}{\sqrt{12}},\\
0 & \text{ for }x\geq\frac{\pi}{\sqrt{12}}.
\end{array}
\right.
\]
In particular, the mean value of $\Delta$ equals $\frac{3}{8}\log3,$ its
variance equals $\left(  \frac{\pi}{8\sqrt{3}}-\frac{9}{64}\left(
\log3\right)  ^{2}\right)  ,$ and the maximal possible sphere packing density
equals $\frac{\pi}{\sqrt{12}}$.
\end{theorem}%

\proof
The lattice $L_{z}$ has density $\Delta$ if and only if $y=y\left(
\Delta\right)  =\frac\pi{4\Delta}.$ The parameter $z=x+iy$ is therefore
uniquely determined by the pair $\left(  x,\Delta\right)  .$ So it can be
taken as the coordinate system on $F.$ We have $dy\left(  \Delta\right)
=-\frac\pi4\frac{d\Delta}{\Delta^{2}}$ . The measure $\mu$ can be rewritten in
coordinates $\left(  x,\Delta\right)  $ as
\[
d\mu=\frac3\pi\frac{dxdy}{y^{2}}=\frac3\pi\frac{dx\left(  \frac\pi
4\frac{d\Delta}{\Delta^{2}}\right)  }{\left(  \frac\pi{4\Delta}\right)  ^{2}%
}=\frac{12}{\pi^{2}}dxd\Delta.
\]
Let $F_{b}=F\cap\left\{  y=b\right\}  .$ The density $p_{\Delta}\left(
a\right)  $ is then equal to $\frac{12}{\pi^{2}}\mathrm{mes}\left(
F_{y\left(  a\right)  }\right)  ,$ where $\mathrm{mes}\left(  \cdot\right)  $
stands for the one-dimensional Lebesgue measure. For $y\left(  a\right)
\ge1,$ i.e., for $a\le\frac\pi4,$ we have $\mathrm{mes}\left(  F_{y\left(
a\right)  }\right)  =1;$ for $y\left(  a\right)  \le\frac{\sqrt{3}}2,$ i.e.,
for $a\ge\frac\pi{2\sqrt{3}},$ we have $\mathrm{mes}\left(  F_{y\left(
a\right)  }\right)  =0.$ For $\frac{\sqrt{3}}2\le y\left(  a\right)  \le1$ the
total length $\mathrm{mes}\left(  F_{y\left(  a\right)  }\right)  $ equals
$2\left(  \frac12-\sqrt{1-y\left(  a\right)  ^{2}}\right)  =1-2\sqrt{1-\left(
\frac\pi{4a}\right)  ^{2}},$ which proves the first statement. The statement
about the maximal value of density follows immediately. The mean value is
given by the elementary integral
\[
\int_{0}^{\infty}x\,p_{\Delta}\left(  x\right)  \,dx=\frac{12}{\pi^{2}}\left(
\int_{0}^{\frac\pi{\sqrt{12}}}x\,\,dx-2\int_{\frac\pi4}^{\frac\pi{\sqrt{12}}%
}x\sqrt{1-\left(  \frac\pi{4x}\right)  ^{2}}\,dx\right)  ,
\]
and a similar equality defines the variance.
\endproof

\section{Random packings}

In this section instead of lattices we will consider configurations $\sigma$
of the point random fields, and we will try to solve for them the same
questions we were discussing above for lattices.

By a point random field in $\mathbb{R}^{n}$ we mean a probability measure
$\mathbb{P} $ on the set $\mathcal{S}$ of all countable locally finite subsets
of $\mathbb{R}^{n}.$ The simplest example of such measure is the Poisson
random field. To define it we first introduce for every $V\subset
\mathbb{R}^{n}$ the notation $\mathcal{S}_{V}$ for the set of all locally
finite subsets of $V,$ and for every $\sigma\in\mathcal{S}$ we denote by
$\sigma_{V}\in\mathcal{S}_{V}$ the intersection $\sigma\cap V.$ We denote by
$\left|  \sigma_{V}\right|  $ the cardinality of the set $\sigma_{V}.$

A random field $\mathbb{P}^{\lambda}$ is called a \textit{Poisson random field
with intensity }$\lambda>0$ if and only if

\noindent$i)$ for every two disjoint subsets $V,W\subset\mathbb{R}^{n},$
$V\cap W=\emptyset,$ the random configurations $\sigma_{V},\sigma_{W}$ are independent;

\noindent$ii)$ for every finite subset $V\subset\mathbb{R}^{n}$ the
conditional distribution of $\sigma_{V}$ under the condition that $\left|
\sigma_{V}\right|  =m$ is just the Lebesgue measure on $V^{m},$ normalized by
the factor $\frac1{\left(  \mathrm{vol}\left(  V\right)  \right)  ^{m}},$
while the probability of the event $\left|  \sigma_{V}\right|  =m$ is given
by
\[
\mathbb{P}\left\{  \sigma:\left|  \sigma_{V}\right|  =m\right\}
=e^{-\lambda\,\mathrm{vol}\left(  V\right)  }\frac{\left(  \lambda
\,\mathrm{vol}\left(  V\right)  \right)  ^{m}}{m!}.
\]
(We recall that the independence property means that for every two events
$A\subset\mathcal{S}_{V},B\subset\mathcal{S}_{W}$ we have $\mathbb{P}\left\{
\sigma:\sigma_{V}\in A,\sigma_{W}\in B\right\}  =\mathbb{P}\left\{
\sigma:\sigma_{V}\in A\right\}  \mathbb{P}\left\{  \sigma:\sigma_{W}\in
B\right\}  $ provided $V,W $ are disjoint.) In what follows we will denote by
$\mathbb{P}$ the Poisson random field with intensity $\lambda=1,$ and we will
omit $\lambda$ from our notation, as well as the adjective ``Poisson''.

{\small More general random fields can be treated by the methods presented
below. These are called Gibbs random fields corresponding to the interactions
}$U.${\small \ Here }$U=U\left(  x,y\right)  ${\small \ is some given function
interpreted as the strength of interaction between two particles situated at
locations }$x\neq y\in\mathbb{R}^{n}.${\small \ A random field }%
$\mathbb{P}^{U}${\small \ is called Gibbs random field with interaction }$U
${\small \ if and only if its conditional distribution inside the finite box
}$V\subset\mathbb{R}^{n},${\small \ given the configuration }$\sigma
_{\mathbb{R}^{n}\backslash V}${\small \ outside it, has density with respect
to the measure }$\mathbb{P}^{\lambda}${\small \ proportional to the ``Gibbs
factor'' }
\[
\exp\left\{  -\sum_{\substack{x\in\sigma_{V}, \\y\in\sigma_{V}\cup
\sigma_{\mathbb{R}^{n}\backslash V} }}U\left(  x,y\right)  \right\}  .
\]
{\small (For a general function }$U${\small \ the Gibbs random field lacks the
independence property }$i)${\small \ above.) To make it into a well-defined
object, one has to put some restrictions on the function }$U.${\small \ The
most general one is called superstability condition. We will not formulate it,
but just note that it is satisfied if }$U${\small \ is repulsive, which just
means that }$U\ge0.${\small \ For further details see \cite{R} or \cite{D}.
The Poisson fields above correspond to zero interaction.}

Let now $\sigma$ be our point random field. Then it is easy to see that for
every $\varepsilon>0$%
\[
\mathbb{P}\left(  \Delta_{N}\left(  \sigma\right)  >\varepsilon\right)
\rightarrow0\text{ as }N\rightarrow\infty.
\]
Since our goal is to construct subsets with positive (and even as big as
possible) geometric density, this is unsatisfactory. To save the situation we
allow to decimate the configuration $\sigma,$ by considering only ``a fraction
$\nu$'' of our collection of points $\sigma,$ where $\nu\in\left(  0,1\right)
.$ To this end we introduce the quantity
\begin{equation}
\Delta_{N,\nu}\left(  \sigma\right)  =\max_{\substack{\tau\subset\sigma_{N}:
\\\nu\left|  \sigma_{N}\right|  -\left|  \sigma_{N}\right|  ^{1/2+\varepsilon
}\le\left|  \tau\right|  \le\nu\left|  \sigma_{N}\right|  +\left|  \sigma
_{N}\right|  ^{1/2+\varepsilon} }}\Delta_{N}\left(  \tau\right)  , \label{31}%
\end{equation}
where $\varepsilon$ is some fixed small number. This fraction is so designed
that for any $\tau$ satisfying the restrictions in (\ref{31}) we have
$\frac{\left|  \tau\right|  }{\left|  \sigma_{N}\right|  }\rightarrow\nu$ as
$N\rightarrow\infty$ for $\sigma$ typical. As the following statement shows,
this improves the situation.

\begin{theorem}
\label{C} With $\mathbb{P}$-probability $1$ the limit
\[
\mathcal{D}\left(  \nu\right)  =\liminf_{N\rightarrow\infty}\Delta_{N,\nu
}\left(  \sigma\right)
\]
does not depend on $\sigma,$ for every $\nu,$ $0<\nu<1.$

$i)$ Let us introduce the function $\nu_{1}\left(  d\right)  =1-\frac{1}%
{2}v_{n}\left(  d\right)  .$ Then for every $d>0$ we have the lower bound
\begin{equation}
\mathcal{D}\left(  \nu_{1}\left(  d\right)  \right)  \geq2^{-n}v_{n}\left(
d\right)  \nu_{1}\left(  d\right)  ,\medskip\medskip\label{13}%
\end{equation}
provided of course that $\nu_{1}\left(  d\right)  >0.$ In particular, for
$d=d_{n},$ where $d_{n}$ satisfies $v_{n}\left(  d_{n}\right)  =1,$ we get
$\nu_{1}\left(  d_{n}\right)  =\frac{1}{2},$ and so
\[
\mathcal{D}\left(  \frac{1}{2}\right)  \geq2^{-n-1}.
\]

$ii)$ Introducing the function $\nu_{2}\left(  d\right)  =1-\frac{1}{2}%
v_{n}\left(  d\right)  e^{-2v_{n}\left(  d\right)  },$ we have for every
$d>0$
\begin{equation}
\mathcal{D}\left(  \nu_{2}\left(  d\right)  \right)  \leq2^{-n}v_{n}\left(
d\right)  \nu_{2}\left(  d\right)  . \label{14}%
\end{equation}

$iii)$ Introducing the function $\nu_{3}\left(  d\right)  =1-\frac{1}{3}%
v_{n}\left(  d\right)  -\frac{1}{6}v_{n}\left(  d\right)  e^{-v_{n}\left(
d\right)  },$ we have for every $d>0$%
\begin{equation}
\mathcal{D}\left(  \nu_{3}\left(  d\right)  \right)  \geq2^{-n}v_{n}\left(
d\right)  \nu_{3}\left(  d\right)  . \label{15}%
\end{equation}
\end{theorem}

\smallskip%
\begin{figure}
[h]
\begin{center}
\includegraphics[
natheight=285.312500pt,
natwidth=285.312500pt,
height=3.9738in,
width=3.9738in
]%
{../../SWP25/docs/TS/ts8.bmp}%
\caption{Upper (3) and lower (2,4) estimates on $\mathcal{D}\left(
\nu\right)  .$}%
\end{center}
\end{figure}

\textbf{Note 1. }The relations (\ref{13})-(\ref{15}) give implicit bounds on
the function $\mathcal{D}\left(  \nu\right)  .$ The relation (\ref{13}) can
easily be rewritten in explicit form: $\mathcal{D}\left(  \nu\right)
\ge2^{-n+1}\nu\left(  1-\nu\right)  .$ The remaining relations can not be
rewritten so easily.

\textbf{Note 2. }The relation (\ref{15}) is an improvement of (\ref{13}), see
Fig. 1.

\textbf{Note 3. }Because of the properties of the function $\nu_{2}\left(
d\right)  $ the relation (\ref{14}) gives an upper estimate on $\mathcal{D}%
\left(  \nu\right)  $ only for the values $\nu\ge1-4e^{-1}\approx0.908\;.$

\textbf{Note 4. }Further improvements of the relations (\ref{13})-(\ref{15})
can also be obtained, see the proof of the theorem below.

\textbf{Note 5. }The above theorem tells us that for almost every
configuration $\sigma$ and for every $N$ large enough we can find a
subconfiguration $\sigma_{N}^{\prime}\subset\sigma_{N}$ with about $\nu\left|
\sigma_{N}\right|  $ points in it and with the geometric density close to
$\mathcal{D}\left(  \nu\right)  .$ However, the configurations $\sigma
_{N}^{\prime}$ might not converge as $N\rightarrow\infty.$

\subsection{Proof of the Theorem 4}

We impose periodic boundary conditions in the box $V_{N},$ which wraps it into
a torus. For a finite subset of points $\sigma\subset V_{N}$, a point $x\in
V_{N}$ and a number $d>0$ we define
\[
m\left(  x,d,\sigma\right)  =\left|  B\left(  x,d\right)  \cap\left\{
\sigma\,\backslash\,x\right\}  \right|  .
\]
This is just the number of points in $\sigma$, different from $x,$ which are
at a distance not bigger than $d$ from $x.$ Let us also introduce the measure
$\delta_{\sigma}$ on $V_{N}$ by putting a $\delta$-measure at each point of
$\sigma:$%
\[
\delta_{\sigma}=\sum_{y\in\sigma}\delta_{y}.
\]
The relevant quantity to look on is now the sum
\[
M\left(  d,\sigma\right)  =\frac12\int_{V_{N}}m\left(  x,d,\sigma\right)
\,\delta_{\sigma}\left(  dx\right)  .
\]
$M\left(  d,\sigma\right)  $\ is the number of pairs of points in $\sigma$\ at
a distance not bigger than $d.$\ The reason we introduce the quantity
$M\left(  d,\sigma\right)  $\ is the following. Let us consider the graph
$G_{d}\left(  \sigma\right)  $\ with vertices at the points of $\sigma$\ and
with edges connecting any two vertices at distance $\le d,$\ then $M\left(
d,\sigma\right)  $\ is precisely the number of edges in $G_{d}\left(
\sigma\right)  .$\ Clearly, in the case $M\left(  d,\sigma\right)  =M,$\ we
can find a subset $\tau\subset\sigma$\ with $\,$%
\[
\left|  \tau\right|  =\left|  \sigma\right|  -M,
\]
such that $d\left(  \tau\right)  \ge d,$\ so the graph $G_{d}\left(
\tau\right)  $\ has no edges and therefore
\begin{equation}
\Delta_{N}\left(  \tau\right)  \ge\dfrac{v_{n}\left(  \frac d2\right)  \left(
\left|  \sigma\right|  -M\right)  }{\left(  2N\right)  ^{n}}. \label{1}%
\end{equation}
Since evidently for $\nu=\dfrac{\left|  \tau\right|  }{\left|  \sigma\right|
}$\ we have
\begin{equation}
\Delta_{N,\nu}\left(  \sigma\right)  \ge\Delta_{N}\left(  \tau\right)  ,
\label{10}%
\end{equation}
the rhs of the estimate (\ref{1}) provides the lower estimate for the quantity
we are interested in.

We introduce now other important quantities, which allow us to obtain better
estimates on the number of points in the subset $\tau\subset\sigma$\ with the
property that $d\left(  \tau\right)  \ge d.$\ Let $M_{1}\left(  d,\sigma
\right)  $\ be the number of isolated edges of the graph $G_{d}\left(
\sigma\right)  .$\ Then for every $\tau$\ with $d\left(  \tau\right)  \ge
d$\ we have
\begin{equation}
\left|  \tau\right|  \le\left|  \sigma\right|  -M_{1}. \label{50}%
\end{equation}
Likewise, we introduce the quantity $M_{3}\left(  d,\sigma\right)  $\ to be
the number of subgraphs of $G_{d}\left(  \sigma\right)  $\ which are maximal
connected components of $G_{d}\left(  \sigma\right)  $\ with 3 vertices and 3
edges (i.e. just isolated triangles), and $M_{2}\left(  d,\sigma\right)  $\ to
be the number of maximal connected components of $G_{d}\left(  \sigma\right)
$\ with 3 vertices and 2 edges. Then we have for all $\tau$\ with $d\left(
\tau\right)  \ge d$\ that
\[
\left|  \tau\right|  \le\left|  \sigma\right|  -M_{1}-M_{2}-2M_{3}.
\]

On the other hand, because of the Theorem \ref{K} below, we can claim the
existence of the subset $\tau\subset\sigma$\ with $d\left(  \tau\right)  \ge
d$\ and such that
\begin{equation}
\left|  \tau\right|  \ge\left|  \sigma\right|  -M_{1}-\frac23\left(
M-M_{1}\right)  \label{51}%
\end{equation}
(see estimate (\ref{32}) below), and even
\[
\left|  \tau\right|  \ge\left|  \sigma\right|  -M_{1}-M_{2}-2M_{3}%
-\frac35\left(  M-M_{1}-2M_{2}-3M_{3}\right)
\]
(see estimate (\ref{33}) below). One can proceed further with such estimates,
introducing the quantities $M_{i}$\ with higher $i$-s; any information we have
on the behavior of the random variables $M_{i}\left(  d,\sigma\right)
$\ provides us with some answer to the question we are interested in.

We start by the study of the random variable $M\left(  d,\sigma_{N}\right)  .$

\begin{lemma}
\label{mean} The mean value
\[
\mathbb{E}\left(  M\left(  d,\sigma_{N}\right)  \right)  =\frac{v_{n}\left(
d\right)  }{2}\left(  2N\right)  ^{n}.
\]
\end{lemma}%

\proof
We first use the identity
\[
\mathbb{E}\left(  M\left(  d,\sigma\right)  \right)  =\mathbb{E}\left[
\mathbb{E}\left(  M\left(  d,\sigma\right)  |\left\{  \left|  \sigma\right|
=k\right\}  \right)  \right]  .
\]
The conditional expectation $\mathbb{E}\left(  M\left(  d,\sigma\right)
|\left\{  \left|  \sigma\right|  =k\right\}  \right)  $\ can be computed in
the following way. Denote by $\chi_{d}\left(  x,y\right)  $\ the function
\[
\chi_{d}\left(  x,y\right)  =\left\{
\begin{array}
[c]{ll}%
1 & \text{ if }\left|  x-y\right|  \le d,\\
0 & \text{ if }\left|  x-y\right|  >d,
\end{array}
\right.
\]
$x,y\in V_{N}.$\ Let $\xi_{1},...,\xi_{k}$\ be $k$\ independent random points
in $V_{N},$\ distributed uniformly according to the Lebesgue measure on
$V_{N}.$\ Then
\[
\mathbb{E}\left(  M\left(  d,\sigma\right)  |\left\{  \left|  \sigma\right|
=k\right\}  \right)  =\mathbb{E}\left(  \sum_{1\le i<j\le k}\chi_{d}\left(
\xi_{i},\xi_{j}\right)  \right)  =\dfrac{k\left(  k-1\right)  }2\dfrac
{v_{n}\left(  d\right)  }{\left(  2N\right)  ^{n}}.
\]
Hence
\[
\mathbb{E}\left(  M\left(  d,\sigma\right)  \right)  =\dfrac{v_{n}\left(
d\right)  }{2\left(  2N\right)  ^{n}}\sum_{k>1}k\left(  k-1\right)
\dfrac{\left(  2N\right)  ^{nk}}{k!}e^{-\left(  2N\right)  ^{n}}\equiv
\dfrac{v_{n}\left(  d\right)  }2\left(  2N\right)  ^{n}.
\]
\endproof

In the same way one can compute the expectation $\mathbb{E}\left(
M_{1}\left(  d,\sigma_{N}\right)  \right)  .$ A particle $x\in\sigma_{N}$
contributes to $M_{1}\left(  d,\sigma_{N}\right)  $ if there is another
particle $y\in\sigma_{N}$ with $\left|  x-y\right|  \le d,$ and there are no
other particles in the union $B\left(  x,d\right)  \cup B\left(  y,d\right)
.$ Since $\mathrm{vol}\left(  B\left(  x,d\right)  \cup B\left(  y,d\right)
\right)  \in\left[  v_{n}\left(  d\right)  ,2v_{n}\left(  d\right)  \right]
,$ we obtain
\[
\frac{v_{n}\left(  d\right)  }2\left(  2N\right)  ^{n}e^{-2v_{n}\left(
d\right)  }\le\mathbb{E}\left(  M_{1}\left(  d,\sigma_{N}\right)  \right)
\le\frac{v_{n}\left(  d\right)  }2\left(  2N\right)  ^{n}e^{-v_{n}\left(
d\right)  }.
\]
\medskip

Before dealing with the variance, we will make a slight generalization.
Namely, let $\phi\left(  x,\sigma\right)  $ be a `local observable', that is,
a function which depends only on the intersection $B\left(  x,\tilde
{d}\right)  \cap\left\{  \sigma\,\backslash\,x\right\}  ,$ for some $\tilde
{d}=\tilde{d}\left(  \phi\right)  .$ We introduce a random variable
$\Phi\left(  \sigma\right)  $ by
\begin{equation}
\Phi\left(  \sigma\right)  =\int_{V_{N}}\phi\left(  x,\sigma_{N}\right)
\,\delta_{\sigma}\left(  dx\right)  . \label{45}%
\end{equation}
For example, the choice $\phi\left(  x,\sigma\right)  =m\left(  x,d,\sigma
\right)  $ corresponds to $\Phi\left(  \sigma\right)  =M\left(  d,\sigma
\right)  .$ In case
\[
\phi\left(  x,\sigma\right)  =\left\{
\begin{array}
[c]{ll}%
1, & \;\text{if }m\left(  x,d,\sigma\right)  =1,\text{ and for the unique }\\
& \;y\in\sigma\text{ with }\left|  x-y\right|  \le d,\,\;m\left(
y,d,\sigma\right)  =1,\\
0 & \;\text{in all other cases}%
\end{array}
\right.
\]
we obtain $\Phi\left(  \sigma\right)  =M_{1}\left(  d,\sigma\right)  .$ We
will impose the following restriction on $\phi:$ there exists a constant
$c>0,$ such that for all $x,y\in\mathbb{R}^{n},$ all $\sigma\subset
\mathbb{R}^{n}$%
\begin{equation}
\left|  \phi\left(  x,\sigma\cup y\right)  -\phi\left(  x,\sigma\right)
\right|  \le c. \label{42}%
\end{equation}
It clearly holds in the above examples, with $c=1$.

\begin{lemma}
The variance $\mathbb{D}\left(  \Phi\left(  \sigma_{N}\right)  \right)  $
satisfies the estimate
\[
\mathbb{D}\left(  \Phi\left(  \sigma_{N}\right)  \right)  \leq C\left(
2N\right)  ^{n},
\]
with $C=C\left(  \phi\right)  .$
\end{lemma}%

\proof
We start with the identity
\begin{equation}
\mathbb{D}\left(  \xi\right)  =\mathbb{D}\left[  \mathbb{E}\left(  \xi
|\eta\right)  \right]  +\mathbb{E}\left[  \mathbb{D}\left(  \xi|\eta\right)
\right]  , \label{2}%
\end{equation}
valid for any two random variables. We are going to apply it for the case of
$\xi=\Phi\left(  \sigma\right)  ,$ while $\eta$ will be the restriction of
$\sigma$ to certain subsets $K\subset V_{N},$ which are called ``corridors''
and which are defined as follows. Consider the partition $\Pi$ of the box
$V_{N}$, formed by the cubic subvolume $V_{l}$ together with all its shifts by
the vectors of the lattice $2l\mathbb{Z}^{n}.$ The number $l$ is chosen to be
equal to $\tilde{d}+D,$ where the number $D$ has to be of the order of
$\tilde{d}:\,D\in\left[  \tilde{d}\,k\left(  n\right)  ,2\tilde{d}\,k\left(
n\right)  \right]  ,$ with the integer $k\left(  n\right)  $ depending only on
the dimension $n.$ For such a partition to exist, we need the ratio $\frac
N{\tilde{d}+D}$ to be an integer; clearly, such a choice of the number $D$ is
possible. The union of the corridors $K_{1}$ is now defined as the $\tilde{d}%
$-neighborhood of the union of the boundaries of all the boxes of $\Pi.$ In
other words, it consists of the strips of the width $2\tilde{d},$ which are
parallel to the various coordinate planes, with the spacing between the two
consecutive one to be equal to $2D.$ (The subscript in the notation $K_{1}$ is
needed due to the fact that later we will have to consider other corridors.)
The reason to introduce the corridors $K_{1}$ is that the random variable
$\Phi\left(  \sigma\right)  ,$ conditioned by the value $\eta$ of the
restriction $\sigma|_{K_{1}},$ turns into the sum of independent random variables.

Let us start with the second term of (\ref{2}), $\mathbb{E}\left[
\mathbb{D}\left(  \xi|\eta\right)  \right]  .$ We first estimate the
contribution to the inner variance, coming from a single box $V_{l}$ of $\Pi.$
That is, we need to consider the box $V_{l-\tilde{d}},$ the Poisson random
field $\sigma$ inside it, together with the fixed configuration $\eta
_{K\left(  V_{l}\right)  } $ in the corridor $K\left(  V_{l}\right)
=V_{l}\,\backslash\,V_{l-\tilde{d}}.$ The random variable we should study, is
the sum
\begin{equation}
\phi^{V_{l}}\left(  \sigma\,|\,\eta_{K\left(  V_{l}\right)  }\right)
=\int_{V_{l-\tilde{d}}}\phi\left(  x,\sigma\right)  \,\delta_{\sigma}\left(
dx\right)  +\int_{V_{l-\tilde{d}}}\left[  \phi\left(  x,\sigma\cup\eta\right)
-\phi\left(  x,\sigma\right)  \right]  \,\delta_{\sigma}\left(  dx\right)  .
\label{41}%
\end{equation}
To estimate its variance we will use the evident relation:
\begin{equation}
\mathbb{D}\left(  \zeta+\chi\right)  =\mathbb{D}\left(  \zeta\right)
+\mathbb{D}\left(  \chi\right)  +2\left(  \mathbb{E}\left(  \zeta\chi\right)
-\mathbb{E}\left(  \zeta\right)  \mathbb{E}\left(  \chi\right)  \right)
\label{3}%
\end{equation}
with $\zeta$ to be the first term in (\ref{41}), and $\chi$ - the second. The
first variance is just a constant:
\[
R_{0}\equiv R_{0}\left(  l,\phi\right)  =\mathbb{D}\left(  \int_{V_{l-\tilde
{d}}}\phi\left(  x,\sigma\right)  \,\delta_{\sigma}\left(  dx\right)  \right)
.
\]
To estimate the covariance term, corresponding to $\mathbb{E}\left(  \zeta
\chi\right)  -\mathbb{E}\left(  \zeta\right)  \mathbb{E}\left(  \chi\right)  $
in (\ref{3}), we rewrite the difference $\left[  \phi\left(  x,\sigma\cup
\eta\right)  -\phi\left(  x,\sigma\right)  \right]  $ as
\[
\phi\left(  x,\sigma\cup\eta\right)  -\phi\left(  x,\sigma\right)  =\sum
_{i=1}^{n}\psi\left(  x,\sigma\cup\eta_{i-1},y_{i}\right)  ,
\]
where
\[
\psi\left(  x,\kappa,y\right)  =\phi\left(  x,\kappa\cup y\right)
-\phi\left(  x,\kappa\right)  .
\]
Here we use some enumeration $\eta_{K\left(  V_{l}\right)  }=\left\{
y_{1},y_{2},...,y_{n}\right\}  ,$ while $\eta_{i}=\left\{  y_{1}%
,y_{2},...,y_{i}\right\}  .$ Then
\begin{align}
&  \mathbb{E}\left(  \zeta\chi\right)  -\mathbb{E}\left(  \zeta\right)
\mathbb{E}\left(  \chi\right)  =\nonumber\\
&  =\sum_{i=1}^{n}{\huge \big\{}\mathbb{E}^{\sigma}\left(  \int_{V_{l-\tilde
{d}}}\phi\left(  x,\sigma\right)  \,\delta_{\sigma}\left(  dx\right)
\int_{V_{l-\tilde{d}}}\psi\left(  x,\sigma\cup\eta_{i-1},y_{i}\right)
\delta_{\sigma}\left(  dx\right)  \right)  -\label{43}\\
&  -\mathbb{E}^{\sigma}\left(  \int_{V_{l-\tilde{d}}}\phi\left(
x,\sigma\right)  \,\delta_{\sigma}\left(  dx\right)  \right)  \mathbb{E}%
^{\sigma}\left(  \int_{V_{l-\tilde{d}}}\psi\left(  x,\sigma\cup\eta
_{i-1},y_{i}\right)  \delta_{\sigma}\left(  dx\right)  \right)  {\huge \big\}%
}.\nonumber
\end{align}
(The symbol $\mathbb{E}^{\sigma}$ means taking expectation in $\sigma;$ $\eta$
is here a fixed parameter.) Note that
\[
\psi\left(  x,\sigma\cup\eta_{i-1},y_{i}\right)  \equiv\phi\left(
x,\sigma\cup\eta_{i-1}\cup y_{i}\right)  -\phi\left(  x,\sigma\cup\eta
_{i-1}\right)  =0
\]
unless $x\in B\left(  y_{i},\tilde{d}\right)  ,$ in which case
\[
\psi\left(  x,\sigma\cup\eta_{i-1},y_{i}\right)  =\psi\left(  x,\sigma
_{B\left(  y_{i},2\tilde{d}\right)  }\cup\eta_{i-1},y_{i}\right)  ,
\]
and
\[
\phi\left(  x,\sigma\right)  \,=\phi\left(  x,\sigma_{B\left(  y_{i}%
,2\tilde{d}\right)  }\right)  .
\]
Therefore
\begin{align*}
&  \mathbb{E}^{\sigma}\left(  \int_{V_{l-\tilde{d}}}\phi\left(  x,\sigma
\right)  \,\delta_{\sigma}\left(  dx\right)  \int_{V_{l-\tilde{d}}}\psi\left(
x,\sigma\cup\eta_{i-1},y_{i}\right)  \,\delta_{\sigma}\left(  dx\right)
\right)  -\\
&  -\mathbb{E}^{\sigma}\left(  \int_{V_{l-\tilde{d}}}\phi\left(
x,\sigma\right)  \,\delta_{\sigma}\left(  dx\right)  \right)  \mathbb{E}%
^{\sigma}\left(  \psi\left(  x,\sigma\cup\eta_{i-1},y_{i}\right)
\,\delta_{\sigma}\left(  dx\right)  \right) \\
&  =\mathbb{E}^{\sigma}\left(  \int_{V_{l-\tilde{d}}}\phi\left(
x,\sigma_{B\left(  y_{i},2\tilde{d}\right)  }\right)  \,\delta_{\sigma}\left(
dx\right)  \int_{V_{l-\tilde{d}}}\psi\left(  x,\sigma_{B\left(  y_{i}%
,2\tilde{d}\right)  }\cup\eta_{i-1},y_{i}\right)  \,\delta_{\sigma}\left(
dx\right)  \right)  -\\
&  -\mathbb{E}^{\sigma}\left(  \int_{V_{l-\tilde{d}}}\phi\left(
x,\sigma_{B\left(  y_{i},2\tilde{d}\right)  }\right)  \,\delta_{\sigma}\left(
dx\right)  \right)  \mathbb{E}^{\sigma}\left(  \int_{V_{l-\tilde{d}}}%
\psi\left(  x,\sigma_{B\left(  y_{i},2\tilde{d}\right)  }\cup\eta_{i-1}%
,y_{i}\right)  \,\delta_{\sigma}\left(  dx\right)  \right)  .
\end{align*}

Because of (\ref{42}), the random variable $\left|  \psi\left(  x,\sigma
\cup\eta_{i-1},y_{i}\right)  \right|  \equiv$\linebreak $\equiv\left|
\phi\left(  x,\sigma\cup\eta_{i-1}\cup y_{i}\right)  -\phi\left(  x,\sigma
\cup\eta_{i-1}\right)  \right|  $ is uniformly bounded by $c\left(
\phi\right)  ,$ while $\left|  \phi\left(  x,\sigma_{B\left(  y_{i},2\tilde
{d}\right)  }\right)  \right|  \le c\left(  \phi\right)  \left|
\sigma_{B\left(  y_{i},2\tilde{d}\right)  }\right|  ,$ so $\mathbb{E}^{\sigma
}\left(  \left|  \phi\left(  x,\sigma_{B\left(  y_{i},2\tilde{d}\right)
}\right)  \right|  \right)  $ is bounded as well. Therefore we can continue in
(\ref{43}) by
\[
\mathbb{E}\left(  \zeta\chi\right)  -\mathbb{E}\left(  \zeta\right)
\mathbb{E}\left(  \chi\right)  \le\frac12R_{1}\left|  \eta_{K\left(
V_{l}\right)  }\right|  ,
\]
where $R_{1}\equiv R_{1}\left(  \phi\right)  ,$ thus having
\begin{align}
\mathbb{D}^{\sigma}\left(  \phi^{V_{l}}\left(  \sigma\,|\,\eta_{K\left(
V_{l}\right)  }\right)  \right)   &  \le R_{0}+R_{1}\left|  \eta_{K\left(
V_{l}\right)  }\right|  +\label{7}\\
&  +\mathbb{D}^{\sigma}\left(  \sum_{i=1}^{n}\left(  \int_{V_{l-\tilde{d}}%
}\psi\left(  x,\sigma_{B\left(  y_{i},2\tilde{d}\right)  }\cup\eta_{i-1}%
,y_{i}\right)  \,\delta_{\sigma}\left(  dx\right)  \right)  \right)
.\nonumber
\end{align}
Two random variables: $A_{i}=\left(  \int_{V_{l-\tilde{d}}}\psi\left(
x,\sigma_{B\left(  y_{i},2\tilde{d}\right)  }\cup\eta_{i-1},y_{i}\right)
\,\delta_{\sigma}\left(  dx\right)  \right)  $ and $\;$\linebreak
$A_{j}=\left(  \int_{V_{l-\tilde{d}}}\psi\left(  x,\sigma_{B\left(
y_{j},2\tilde{d}\right)  }\cup\eta_{j-1},y_{j}\right)  \,\delta_{\sigma
}\left(  dx\right)  \right)  $ - are independent, provided $\left|
y_{i}-y_{j}\right|  >2\tilde{d}.$ Hence, applying again (\ref{3}),
\begin{align}
&  \mathbb{D}^{\sigma}\left(  \sum_{i=1}^{n}\left(  \int_{V_{l-\tilde{d}}}%
\psi\left(  x,\sigma_{B\left(  y_{i},2\tilde{d}\right)  }\cup\eta_{i-1}%
,y_{i}\right)  \,\delta_{\sigma}\left(  dx\right)  \right)  \right)
=\mathbb{D}^{\sigma}\left(  \sum_{i=1}^{n}A_{i}\right) \label{444}\\
&  =\sum_{i=1}^{n}\mathbb{D}^{\sigma}A_{i}+2\sum_{\substack{i\neq j \\\left|
y_{i}-y_{j}\right|  <2\tilde{d} }}\left[  \mathbb{E}\left(  A_{i}A_{j}\right)
-\mathbb{E}\left(  A_{i}\right)  \mathbb{E}\left(  A_{j}\right)  \right]
.\nonumber
\end{align}
The first term in (\ref{444}) can be estimated from above by
\begin{equation}
R_{2}\left(  \phi\right)  \left|  \eta_{K\left(  V_{l}\right)  }\right|  ,
\label{5}%
\end{equation}
while the second term -- by
\begin{equation}
R_{3}\left(  \phi\right)  \int_{K\left(  V_{l}\right)  }m\left(  y,2\tilde
{d},\eta\right)  \,\delta_{\eta}\left(  dy\right)  . \label{6}%
\end{equation}
Putting together (\ref{7}), (\ref{444}), (\ref{5}) and (\ref{6}), we get
\begin{align*}
\mathbb{D}\left(  \phi^{V_{l}}\left(  \sigma\,|\,\eta_{K\left(  V_{l}\right)
}\right)  \right)   &  \le R_{0}+R_{4}\left|  \eta_{K\left(  V_{l}\right)
}\right|  +\\
&  +R_{3}\int_{K\left(  V_{l}\right)  }m\left(  y,2\tilde{d},\eta\right)
\,\delta_{\eta}\left(  dy\right)  .
\end{align*}
Hence the total variance
\begin{align*}
\mathbb{D}\left(  \xi|\eta\right)   &  \equiv\sum_{V\in\Pi}\mathbb{D}\left(
\phi^{V}\left(  \sigma_{V}\,|\,\eta_{K\left(  V\right)  }\right)  \right)
\le\\
&  \le\left(  \dfrac N{\tilde{d}+D}\right)  ^{n}R_{0}+R_{4}\int_{K_{1}}%
\delta_{\sigma}\left(  dy\right)  +R_{3}\int_{K_{1}}m\left(  y,2\tilde
{d},\sigma\right)  \,\delta_{\sigma}\left(  dy\right) \\
&  <\left(  \dfrac N{\tilde{d}+D}\right)  ^{n}R_{0}+R_{4}\int_{V_{N}}%
\delta_{\sigma}\left(  dx\right)  +R_{3}\int_{V_{N}}m\left(  x,2\tilde
{d},\sigma\right)  \,\delta_{\sigma}\left(  dx\right)  ,
\end{align*}
and the expectation $\mathbb{E}\left[  \mathbb{D}\left(  \xi|\eta\right)
\right]  $ is indeed less than $C_{1}\left(  2N\right)  ^{n}$ for some
$C_{1},$ according to the Lemma \ref{mean}.

Going now to the first term, $\mathbb{D}\left[  \mathbb{E}\left(  \xi
|\eta\right)  \right]  ,$ in (\ref{2}), we will slightly abuse the notation,
denoting by $\sigma$ the Poisson field outside the corridor $K_{1},$ with
$\eta$ denoting the Poisson field on $K_{1}.$ We have
\begin{align*}
\mathbb{E}^{\sigma}\left(  \Phi\left(  \sigma\cup\eta\right)  |\eta\right)
&  =\left(  \dfrac N{\tilde{d}+D}\right)  ^{n}\mathbb{E}^{\sigma}\left(
\int_{V_{l-\tilde{d}}}\phi\left(  x,\sigma\right)  \,\delta_{\sigma}\left(
dx\right)  \right)  +\\
&  \mathbb{E}^{\sigma}\left(  \int_{V_{N}}\left[  \phi\left(  x,\sigma\cup
\eta\right)  -\phi\left(  x,\sigma\right)  \right]  \,\delta_{\sigma\cup\eta
}\left(  dx\right)  \right)  .
\end{align*}
The first expectation is just a constant:
\[
R_{5}\left(  \phi\right)  =\mathbb{E}\left(  \int_{V_{l-\tilde{d}}}\phi\left(
x,\sigma\right)  \,\delta_{\sigma}\left(  dx\right)  \right)  ,
\]
and so does not contribute to the variance. Therefore we need to study the
variance of the random variable
\begin{equation}
\Phi^{1}\left(  \eta\right)  =\mathbb{E}^{\sigma}\left(  \int_{V_{N}}\left[
\phi\left(  x,\sigma\cup\eta\right)  -\phi\left(  x,\sigma\right)  \right]
\,\delta_{\sigma\cup\eta}\left(  dx\right)  \right)  . \label{46}%
\end{equation}
We want to argue that this function can be represented similarly to
(\ref{45}):
\begin{equation}
\Phi^{1}\left(  \eta\right)  =\int_{K_{1}}\phi^{1}\left(  y,\eta\right)
\,\delta_{\eta}\left(  dy\right)  , \label{47}%
\end{equation}
with the function $\phi^{1}\left(  x,\eta\right)  $ satisfying the analog of
relation (\ref{42}). To show this let us first fix an enumeration $\eta
_{K_{1}}=\left\{  y_{1},y_{2},...,y_{n}\right\}  $ for every configuration
$\eta_{K_{1}}.$ (This even can be done in a measurable way, but since this
enumeration will not be needed in the final analysis, we will not elaborate on
this point.) Define $\eta_{i}=\left\{  y_{1},y_{2},...,y_{i}\right\}  $ and
rewrite (\ref{46}) as
\begin{align*}
\Phi^{1}\left(  \eta\right)   &  =\sum_{i=1}^{n}\mathbb{E}^{\sigma}%
{\huge \big[}\int_{V_{N}}\phi\left(  x,\sigma\cup\eta_{i}\right)
\delta_{\sigma\cup\eta_{i}}\left(  dx\right)  -\\
&  -\int_{V_{N}}\phi\left(  x,\sigma\cup\eta_{i-1}\right)  \delta_{\sigma
\cup\eta_{i-1}}\left(  dx\right)  {\huge \big].}%
\end{align*}
We define now the function
\begin{align}
&  \tilde{\phi}^{1}\left(  y,\eta\right)  =\mathbb{E}^{\sigma}{\huge \big[%
}\phi\left(  y,\sigma\cup\eta_{i-1}\cup y\right)  +\label{48}\\
&  \left(  \int_{V_{N}}\phi\left(  x,\sigma\cup\eta_{i-1}\cup y\right)
-\phi\left(  x,\sigma\cup\eta_{i-1}\right)  \right)  \delta_{\sigma\cup
\eta_{i-1}}\left(  dx\right)  {\huge \big].}\nonumber
\end{align}
It is almost what we need. Indeed, the relation (\ref{47}) is straightforward,
while the estimate (\ref{42}) follows from the same estimate for $\phi.$ The
only drawback is that the function $\tilde{\phi}^{1}\left(  y,\eta\right)  $
depends not only on $\eta,$ but also on the ordering we choose. However, if we
now pass from $\tilde{\phi}^{1}\left(  y,\eta\right)  $ to its symmetrization
$\phi^{1}\left(  y,\eta\right)  $, obtained by averaging over all possible
orderings on $\eta$, we get what we want.

\smallskip The net result of the above discussion is that the estimate of the
variance of the random variable $\Phi\left(  \sigma_{N}\right)  $ is reduced
with the help of the identity (\ref{2}) to the estimate of the variance of the
random variable $\Phi^{1}\left(  \eta\right)  .$ The key advantage of passing
from $\Phi$ to $\Phi^{1}$ is that the latter random variable depends only on
restriction $\eta$ of $\sigma$ to ``$\left(  n-1\right)  $-dimensional''
subset $K_{1}$ of $\mathbb{R}^{n}.$ For example, in case $n=1$ the set $K_{1}$
splits into union of disjoint segments, and the random variable $\Phi
^{1}\left(  \eta\right)  $ is then the sum of independent random variables,
corresponding to these segments. Hence, the variance of this sum is just the
sum of the variances. For general values of $n$ we have to repeat the above
scheme for $\left(  n-1\right)  $ more times, reducing after each step the
``dimension'' of the corridors by one.

So we proceed as follows. To estimate the variance of $\Phi^{1}\left(
\eta\right)  $ we again apply the identity (\ref{2}), with the following
choices: $\xi=\Phi^{1}\left(  \sigma|_{K_{1}}\right)  ,$ while $\eta$ is the
restriction of $\sigma$ to the ``$\left(  n-2\right)  $-dimensional'' corridor
$K_{2}\subset K_{1},$ defined by the following construction. Let $w_{1}\subset
V_{l}$ be the subset of $2n$ points of the cube $V_{l},$ consisting of the
centers of all its $\left(  n-1\right)  $-dimensional faces. (The previous
subset $w_{0}\subset V_{l}$ contained just one point: the center of $V_{l} $
itself.) Then
\[
K_{2}=K_{1}\,\backslash\,\bigcup_{b\in\left\{  w_{1}+2l\mathbb{Z}^{d}\right\}
}\left(  V_{l-2\tilde{d}}+b\right)  .
\]
(Again, for $n=2$ the set $K_{2}$ splits into disjoint union.) The key feature
of thus defined corridor is that the random variable $\Phi^{1}\left(
\sigma|_{K_{1}}\right)  $ conditioned by the value of the restriction
$\sigma|_{K_{2}}$ splits into the sum of independent random variables, so we
can repeat the above arguments. The rest of the proof is just this repetition
and is omitted.
\endproof

Let $\xi$ be any random variable. The \textit{Chebychev inequality} claims
that for any $\varepsilon>0$
\[
\mathbb{P}\left\{  \left|  \xi-\mathbb{E}\left(  \xi\right)  \right|
\ge\varepsilon\right\}  \le\frac{\mathbb{D}\left(  \xi\right)  }%
{\varepsilon^{2}}.
\]
Applying it to $M\left(  d,\sigma_{N}\right)  $ we get the following

\begin{proposition}
Let $\varepsilon>0$ and $\delta>1/2$ be fixed. Then
\[
\mathbb{P}\left\{  \left|  \frac{M\left(  d,\sigma_{N}\right)  -\frac
{v_{n}\left(  d\right)  }{2}\left(  2N\right)  ^{n}}{\left(  \left(
2N\right)  ^{n}\right)  ^{\delta}}\right|  >\varepsilon\right\}  \rightarrow0
\]
as $N\rightarrow\infty.$
\endproof
\medskip
\end{proposition}

\medskip\textbf{End of the proof of Theorem \ref{A}. }Let us fix some value of
the distance $d.$ Let $N$ be large enough, $\varepsilon$ be fixed, and
$\sigma_{N}$ be a typical configuration in $V_{N}.$ Then it follows from the
last proposition that with probability $\mathbb{P}$ going to $1$ as
$N\rightarrow\infty$ the configuration $\sigma_{N}$ has the properties:
\begin{equation}
\left|  \frac{M\left(  d,\sigma_{N}\right)  -\frac{v_{n}\left(  d\right)
}2\left(  2N\right)  ^{n}}{\left(  \left(  2N\right)  ^{n}\right)  ^{2/3}%
}\right|  <\varepsilon\label{11}%
\end{equation}
and
\begin{equation}
\left|  \frac{\left|  \sigma_{N}\right|  -\left(  2N\right)  ^{n}}{\left(
\left(  2N\right)  ^{n}\right)  ^{2/3}}\right|  <\varepsilon\label{12}%
\end{equation}
(again by Chebychev).

Plugging in these data into estimates (\ref{1}), (\ref{10}), we find that for
$\nu=1-\frac12v_{n}\left(  d\right)  $%
\[
\Delta_{N,\nu}\left(  \sigma\right)  \ge v_{n}\left(  \frac d2\right)  \left(
1-\frac12v_{n}\left(  d\right)  \right)  +o\left(  N^{-1/4}\right)
\]
uniformly over configurations $\sigma_{N},$ satisfying (\ref{11}), (\ref{12}),
which proves (\ref{13}). The proof of the rest of the Theorem follows by
similar arguments, applied to relations (\ref{50}), (\ref{51}).
\endproof

\subsection{Vertex covering number of a graph}

In this subsection we prove a result from the graph theory, which was used
above. The Theorem \ref{K} below contains in fact a result slightly stronger
than what was needed.

Let $G=\left(  V\left(  G\right)  ,E\left(  G\right)  \right)  $ be a finite
connected graph without loops and multiple edges. Let $v\left(  G\right)
=\left|  V\left(  G\right)  \right|  $ be the number of its vertices and
$e\left(  G\right)  =\left|  E\left(  G\right)  \right|  $ the number of its
edges. A set $A\subseteq V\left(  G\right)  $ is called a \textit{covering
vertex set} if any vertex of $G$ is a neighbor of a vertex of $A$. Another way
to look at this property of $A$ is to say that if we cross out all the
vertices of $A $ together with all edges having at least one end in $A$, no
edges are left. The \textit{vertex covering number} $\alpha\left(  G\right)  $
is defined as the smallest number of elements in a vertex covering set.

\begin{theorem}
\label{K} Let $G$ be as above, with $e\left(  G\right)  >0,$ then
\[
\alpha\left(  G\right)  \leq\frac{e\left(  G\right)  +1}{2}\,,
\]
and, in particular, for $e\left(  G\right)  $ even
\[
\alpha\left(  G\right)  \leq\frac{1}{2}e\left(  G\right)  \,.
\]
Therefore, if $e\left(  G\right)  \geq2,$ then
\begin{equation}
\alpha\left(  G\right)  \leq\frac{2}{3}e\left(  G\right)  \,, \label{32}%
\end{equation}
and if $e\left(  G\right)  \geq4,$ then
\begin{equation}
\alpha\left(  G\right)  \leq\frac{3}{5}e\left(  G\right)  \,. \label{33}%
\end{equation}
Also,
\begin{equation}
\max_{G,v\left(  G\right)  =v}\,\,\frac{\alpha\left(  G\right)  }{e\left(
G\right)  }=\left\{
\begin{array}
[c]{ll}%
\frac{1}{2}+\frac{1}{2v}\medskip & \text{ for}\mathrm{\,\,}v\,\,\text{odd,}\\
\frac{1}{2}+\frac{1}{2\left(  v-1\right)  } & \text{ for}\mathrm{\,\,}%
v\,\,\text{even,}%
\end{array}
\right.  \label{35}%
\end{equation}
where the maximum is taken over all graphs $G$ of the described type such that
$v\left(  G\right)  =v$.
\end{theorem}

We say that a vertex is an \textit{end vertex }if there is exactly one edge
adjacent to it.

\begin{lemma}
\label{L} Let $G$ be as above, with $v\left(  G\right)  \geq3.$ Then there
exists a non-end vertex such that if we cross it out together with all its
edges, and then cross out all resulted isolated vertices, the resulting graph
will be connected.
\end{lemma}%

\proof
Suppose the assertion to be false. Denote by $V_{ne}\left(  G\right)  $ the
set of all non-end vertices of $G.$ Then for every vertex $s\in V_{ne}\left(
G\right)  $ the result of crossing it out together with all its edges,
followed by crossing out all isolated vertices, is a graph $G_{s}$ with
several components: $G_{s}=G_{s,1}\cup G_{s,2}\cup\dots\cup G_{s,m_{s}}$ with
$m_{s}\ge2$. Let $e_{\min}\left(  s\right)  $ be the minimum number of edges
in $G_{s,i},$ $i=1,2,...,m_{v}$. Define $s_{0}\in V_{ne}\left(  G\right)  $ to
be a minimizer of the function $e_{\min}\left(  \cdot\right)  .$ Fix a
component $G_{s_{0}}^{0}$ with $e_{\min}\left(  s_{0}\right)  $ edges. Note
that $G-G_{s_{0}}^{0}$ is connected. Take any vertex $v_{1}\in G_{s_{0}}^{0}$
which is a neighbor of $s_{0}. $ Clearly, $s_{1}\in V_{ne}\left(  G\right)  $.
Let us cross it out from $G$. By assumption, $G_{s_{1}}$ also has several
components, the one containing $G-G_{s_{0}}^{0}$ is not the smallest one,
hence the smallest one lies in $G_{s_{0}}^{0}$ and has therefore less edges
than $e_{\min}\left(  s_{0}\right)  $, which brings us to contradiction.
\endproof

\textbf{Proof of the Theorem \ref{K}. }We argue by induction on the number of
edges, and we use the notation of the proof of the Lemma \ref{L} above. The
case of one edge is clear. By Lemma \ref{L} we can find a non-end vertex $s\in
V_{ne}\left(  G\right)  $, such that the cross-out graph $G_{s}$ remains
connected; since $\alpha\left(  G\right)  \le\alpha\left(  G_{s}\right)  +1$
and $e\left(  G_{s}\right)  \le e\left(  G\right)  -2,$ we get the first four statements.

To prove (\ref{35}) we first prove the upper bound. We have: $\frac
{\alpha\left(  G\right)  }{e\left(  G\right)  }\le\frac12+\frac1{2e\left(
G\right)  }$. Note that always $e\left(  G\right)  \ge v\left(  G\right)  -1$,
therefore $\frac{\alpha\left(  G\right)  }{e\left(  G\right)  }\le
\frac12+\frac1{2\left(  v\left(  G\right)  -1\right)  }$, while $e\left(
G\right)  \ge v\left(  G\right)  $ and $\frac{\alpha\left(  G\right)
}{e\left(  G\right)  }\le\frac12+\frac1{2v\left(  G\right)  }$ whenever there
is a cycle in $G$. To prove that the left hand side of (\ref{35}) is less than
or equal to the right hand side, the only case left is when $G$ is a tree with
$v\left(  G\right)  $ odd. This is done by induction on $v$. For $v=3$ the
statement is clear. If $v>3$, fix a vertex $s_{0}\in V\left(  G\right)  $ and
take an end vertex $s_{1}$ most distant from $s_{0}$ . The only vertex $s_{2}$
adjoint to it has the graph $G_{s_{2}}$ connected. We have $\alpha\left(
G\right)  \le\alpha\left(  G_{s_{2}}\right)  +1$, $e\left(  G\right)
=v\left(  G\right)  -1$ and $e\left(  G_{s_{2}}\right)  =v\left(  G_{s_{2}%
}\right)  -1$ (since both $G$ and $G_{s_{2}}$ are trees). If $v\left(
G_{s_{2}}\right)  =v\left(  G\right)  -2,$ then by the induction hypothesis
\begin{align*}
\alpha\left(  G\right)   &  \le\alpha\left(  G_{s_{2}}\right)  +1\le
\frac{e\left(  G_{s_{2}}\right)  }2+\frac{e\left(  G_{s_{2}}\right)
}{2v\left(  G_{s_{2}}\right)  }+1\\
&  =\frac12\left(  v\left(  G_{s_{2}}\right)  -1+\frac{v\left(  G_{s_{2}%
}\right)  -1}{v\left(  G_{s_{2}}\right)  }+2\right) \\
&  =\frac12\left(  v\left(  G_{s_{2}}\right)  -\frac1{v\left(  G_{s_{2}%
}\right)  }+2\right) \\
&  \le\frac12\left(  v\left(  G\right)  -\frac1{v\left(  G\right)  }\right)
=\frac{e\left(  G\right)  }2\left(  1+\frac1{v\left(  G\right)  }\right)  \,.
\end{align*}
If $v\left(  G_{s_{2}}\right)  \le v\left(  G\right)  -3,$ then similarly
\begin{align*}
\alpha\left(  G\right)   &  \le\alpha\left(  G_{s_{2}}\right)  +1\le
\frac{e\left(  G_{s_{2}}\right)  }2+\frac{e\left(  G_{s_{2}}\right)
}{2\left(  v\left(  G_{s_{2}}\right)  -1\right)  }+1\\
&  =\frac12\left(  v\left(  G_{s_{2}}\right)  +2\right) \\
&  \le\frac12\left(  v\left(  G\right)  -\frac1{v\left(  G\right)  }\right)
=\frac{e\left(  G\right)  }2\left(  1+\frac1{v\left(  G\right)  }\right)  \,.
\end{align*}

Since the estimate (\ref{35}) is achieved on the line graph (for $v$ even) and
on the circle graph (for $v$ odd), the proof is complete.
\endproof
\medskip

\textbf{Acknowledgments.} We would like to thank V. Kleptsyn, whose proof of
Theorem \ref{K} is reproduced above, V. Dolotin for his help with Fig. 1 and
L.Bassalygo for valuable remarks.


\begin{thebibliography}{Oe1}
\bibitem[CS]{CS}J. Conway, N.J.A. Sloane. \textit{Sphere Packings, Lattices
and Groups.} Springer Verlag, 1988.

\bibitem[D]{D}R.L. Dobrushin. \textit{Gibbsian random fields for particles
without hard core}. Theor. Mat. Physics, \textbf{4}, N1, 101-118, 1970.

\bibitem[H]{H}E. Hlawka. \textit{Zur Geometrie der Zahlen}, Math. Zeit.,
\textbf{49, }285-312 (1944).

\bibitem[M]{M}H. Minkovski. \textit{Volumen und Oberfl\"{a}che}, Mathematische
Ann., \textbf{57, }447-495 (1903), see also in: H. Minkovski.
\textit{Abhandlungen}, Chelsea Publ. Company, NY, v.2, 230-276 (1967).

\bibitem[Oe1]{Oe1}J. Oesterl\'{e}. \textit{Empilements de sph\`{e}res.} Sem.
Bourbaki, 1989-1990, exp.727. Ast\'{e}risque 189-190, 375-397.

\bibitem[Oe2]{Oe2}J. Oesterl\'{e}. \textit{Densit\'{e} maximale des
empilements de sph\`{e}res en dimension 3 [d'apr\`{e}s Thomas C. Hales et
Samuel P. Ferguson].} S\'{e}m. Bourbaki, 1998-1999, exp.863.

\bibitem[RT]{RT}M.Yu. Rosenbloom, M.A. Tsfasman. \textit{Multiplicative
lattices in global fields}. Invent. Math., \textbf{101}, 687-696, 1990.

\bibitem[R]{R}D. Ruelle. \textit{Statistical mechanics: Rigorous results.} W.
A. Benjamin, Inc., New York-Amsterdam, 1969.

\bibitem[Sch]{Sch}W. Schmidt. \textit{The measure of the set of admissible
lattices.} Proc. Amer. Math. Soc., \textbf{9,} 390--403, 1958.

\bibitem[TV]{TV}M.A. Tsfasman, S.G.Vl\u{a}du\c{t}. \textit{Algebraic-Geometric
Codes.} Kluwer Acad. Publ., 1991.
\end{thebibliography}
\end{document}